\documentclass[twocolumn,showpacs,preprintnumbers,floatfix,amssymb,prb]{revtex4}
\usepackage{graphicx}
\usepackage{amsmath}
\usepackage{dcolumn}
\usepackage{bm}
\newcommand{\TMTSF}{(TMTSF)$_2$PF$_6$~}
\newcommand{\TMTSFC}{(TMTSF)$_2$ClO$_4$~}
\newcommand{\Hamil}{{\cal H}}

\newcommand{\rr}{{\mathbf r}}

\begin{document}

\title{Magic angle effects in the interlayer magnetoresistance of 
       quasi-one-dimensional metals due to interchain incoherence}

\author{Urban Lundin}
\email{lundin@physics.uq.edu.au}
\author{Ross H.\ McKenzie}
\affiliation{Department of Physics, University of Queensland,
             Brisbane Qld 4072, Australia}

\date{\today}

\begin{abstract}
The dependence of the 
magnetoresistance of quasi-one-dimensional metals on the direction of the 
magnetic field 
show dips when the field is tilted at the so called magic angles 
determined by the structural dimensions of the materials. 
There is currently no accepted explanation 
for these magic angle effects. We present a possible explanation.  
Our model is based on the assumption that, the intralayer transport in 
the second most conducting direction has a small contribution from 
incoherent electrons. This incoherence is modelled by a small uncertainty in 
momentum 
perpendicular to the most conducting (chain) direction. 
Our model predicts the magic angles seen in interlayer 
transport measurements for different orientations of the field.
We compare our results to predictions by other models and to experiment. 
\end{abstract}
\maketitle


\section{Introduction}
There is a fundamental relation between quantum coherence of electronic 
properties and transport properties in strongly correlated 
metals~\cite{lee96,clarke98,valla02,lundin03,biermann01}. 
Scattering of electrons 
affects the transport, but also blurs information about the momentum of the 
electron, and therefore changes the coherence of the electrons, or 
quasi-particle excitations. 
Generally, the effect of strong electronic correlations and incoherent 
excitations are enhanced in systems of reduced dimensionality. A striking 
example of this are Luttinger liquids in one dimension. 
The quasi one-dimensional Bechgaard salts 
(TMTST)$_2$X (X=PF$_6$, ClO$_4$, NO$_3$, $\ldots$) 
show a rich phase diagram ranging from field induced spin 
density waves to insulators and superconductors, depending 
on pressure and anion~\cite{ishiguro,chaikin96}. 
The structures are highly anisotropic, and show 
interesting features as a magnetic field is applied. 

Lebed~\cite{lebed86} 
predicted that resistance maxima would occur when orbits along 
directions the crystal are commensurate with the applied field at 
the so called ''magic angles'' (MA) where $\tan\theta=lb/c$, where 
$\theta$ is the angle between the magnetic filed, tilted in the $(y,z)$-plane, 
and the least conducting direction, $z$, 
$b$ and $c$ are lattice constants, and $l$ is an integer. 
The MA were later discovered~\cite{naughton91}, but not as maxima but as dips 
in the angular dependence of the magnetoresistance. 
MA effects are also seen~\cite{biskup00} in the (DMET-TSeF)$_2$X 
family where X=AuCl$_2$, AuI$_2$.  
The theory was later modified to explain why dips should be 
found~\cite{lebed94}. 
The idea presented is that periodic motion is induced at the MA and, 
provided there is even a small overlap in the direction of the applied field, 
the electron-electron interaction becomes more two dimensional which would 
produce a dip in the magnetoresistance (MR) at the MA. 
Alternative ideas and explanations have since then appeared in the 
literature.~\cite{osada92,maki92,osada02,lebed03,lebed04,mckenzie00,moses00}  
Most of them are based on a small overlap of electronic wave-functions 
in the direction of the 
magnetic field, but suggestions based on a 
Luttinger liquid approach have also been made~\cite{strong94,georges00}.  
The theory developed by Osada, Kagoshima and Miura~\cite{osada92} 
captures many details of the 
experimental data~\cite{chashechkina01}. However, that theory 
requires the existence of very long-range hopping (for example, the second 
nearest neighbor hopping integral in a tight-binding model is of the same 
order as the next-nearest neighbor integral). 
Further, if $B_x=0$, $x$ being the most conducting direction, i.e., along the 
one-dimensional chain of molecules, 
the theory predicts that there would be no MA seen in the interlayer 
conductivity $\sigma_{zz}$. Further, it 
does not explain the dip in the MR when the magnetic field lies in the plane 
($\theta=90^\circ$). The data for in-plane magnetic field is affected by the 
fact that the sample is superconducting and the upper critical field 
H$_{c_2}$ for an in-plane magnetic 
field is quite large~\cite{naughton97}. 
The consensus is that there is no accepted theory behind the 
appearance of the MA. 
The experimental situation is also unclear at the moment. 
Some groups report  that the MR has the same behavior in 
\textit{all} directions of the 
current~\cite{kang92,chashechkina98,chashechkina01}. 
Whereas other experiments disagree~\cite{kang03},
and claim that, due to defects in the crystals, the electrons are forced to 
travel in one direction past the defects, so that 
there can be a contribution 
from, e.g., the resistivity in the $x$-direction to the resistivity in the 
$z$-direction, producing an apparently similar angular behaviour of 
$\rho_{xx}$ and $\rho_{zz}$. 

The MA effects are also seen in torque measurements, as 
measured by Naughton {\it et al.}~\cite{naughton91}. 
This has been discussed theoretically by Yakovenko~\cite{yakovenko92}.
Since the torque can be related to the free energy, 
it is likely that MA effects reflects the ground state electronic 
properties of the material. Further, a big Nernst signal has been detected at 
the MA~\cite{wu03}, and has been discussed 
theoretically~\cite{ong04}.  

The crystal is oriented so that $x$ is the most conducting 
direction, followed by the $y$-direction. The ($x,y$)-plane defines the 
layered structure. Typically, the hopping in the three different directions 
are estimated to be of the order~\cite{gorkov96,chaikin96} 
$t_x:t_y:t_z \sim 2000 K : 200 K : 10 K $. 


Here we present an alternative explanation of the occurrence of the MA. 
Our physical picture is the following. 
The strongly anisotropic structure of the material affects the coherence 
of the particles in the crystal, as well as increasing the effect of 
the electron correlations~\cite{vescoli98,lorentz02}. 
In a previous paper, we discussed 
a model for transport in layered materials based on a coherence-incoherence 
crossover as a function of temperature~\cite{lundin03}. 
Along the $x$-direction the motion is assumed to be coherent. 
The least conducting $z$-direction is assumed to be incoherent, and 
in the $y$-direction, the motion is predominantly coherent, but 
with a small incoherent contribution. 
We will show that 
the loss of coherence in the $y$-direction is directly responsible for the 
MA in the conductance measured along the $z$-direction. 
Even a small amount of incoherence gives rise to a sizable 
effect seen at the MA. 

\section{Model}
We modell the system as a quasi-one-dimensional metal. 
We introduce coordinates such that $a$ is the lattice spacing 
in the $\hat{x}$-direction, 
$b$ in the $\hat{y}$, and $c$ in the $\hat{z}$, 
the layers lie in the $(x,y)$-plane. 
Due to the layered structure, the Hamiltonian is divided into 
intralayer and interlayer contributions 
\begin{equation}
\Hamil=\Hamil_{\parallel}+\Hamil_{\perp},
\end{equation}
where $\Hamil_{\parallel}$ describes the 2D $(x,y)$-layer and includes all 
many-body interactions within each layer, and 
$\Hamil_{\perp}=t_{\perp}\sum_{<i,j>} (c^{\dagger}_ic_j + \mathrm{h.c.})$ 
describes the tunneling between nearest neighbors 
in the $z$-direction. 
Because of the layered crystal structure, we assume that Coulomb correlations 
between the layers are small, and the separation is valid. 
Later, we will further specify $\Hamil_{\parallel}$ for 
quasi-one-dimensional systems. 
If we have a magnetic field in the $(y,z)$-plane 
the vector potential, $\vec{A}$, for the
magnetic field $\vec{B} = (0,B_y,B_z)= (0, B\sin\theta , B\cos\theta)$. 
In the Landau gauge $\vec{A}$ is
\begin{equation}
\vec{A} = (zB_y,xB_z,0). \nonumber
\end{equation}
We are going to study transport in the $z$-direction, i.e., transport between 
the anisotropic two-dimensional layers. Let us consider two adjacent layers. 
The vector potential in the two layers 
are not equal but differ by a gauge transformation 
$\vec{A}_{2}=\vec{A}_{1}+ \vec{\nabla} \Lambda$, $1$ and $2$ indicate 
the layer, and $\Lambda=cB_yx$. 
At small bias we can use linear response theory to calculate the 
current between the layers. 
At low temperatures, 
only electrons at the Fermi-energy contribute to the conductivity in the least 
conducting direction, and it can be written as a function of only the in-plane 
Green functions~\cite{moses99} due to the separation of intralayer and 
interlayer contributions in the Hamiltonian. Separating the 
current-current correlation function we get that the conductivity 
is given by~\cite{moses99} 
\begin{eqnarray}
&&\sigma_{zz}=
\nonumber \\ 
&&\frac{e^2t_{\perp}^2c}{\hbar\pi L_xL_y}
\int {\mathrm d}\rr \int {\mathrm d}\rr'
\left[G^{1+}(\rr,\rr',E_F)G^{2-}(\rr',\rr,E_F) \right.
\nonumber \\ && \left. +
                G^{1-}(\rr',\rr,E_F)G^{2+}(\rr,\rr',E_F)\right], 
\nonumber \\
\label{conduct:eq}
\end{eqnarray}
where $G^{1+}(\rr,\rr',E_F)$ denotes the electronic Green function (GF) 
within a single layer. 
Here, $L_xL_y$ are the dimensions of the sample in the $x$-, and $y$-direction
respectively.  There is an 
indirect dependence on the distance between the layers in $t_{\perp}$. 
We stress that this is a very general expression and contains all the 
many-body effects within each layer. 

\subsection{Non-interacting Green function for quasi-one-dimensional materials
            in a magnetic field}
Let us now look at the Hamiltonian in the absence of electron-electron 
interactions. We assume that the spectra in the most conducting direction 
can be linearized. Then, the Hamiltonian for a layer in a tilted magnetic 
field is (see Ref.~\onlinecite{moses99}) 
\begin{equation}
\Hamil_{\parallel}^0 = \alpha  v_F (-i\hbar \partial_x + ezB\sin\theta )-
2 t_y \cos\left[b \left(-i\hbar
\partial_y - e x B \cos\theta \right)\right],
\end{equation}
where $\alpha =\pm 1$ denotes which sheet of the Fermi surface the electron 
is on. $v_F$ is the Fermi velocity, and $t_y$ is the interchain 
hopping-integral. 
The wavefunction is written as
\begin{equation}
\psi(x,y,t)=\exp\left\{ i\left[\frac{-\epsilon t}{\hbar} +k_xx+k_yy 
                        -\alpha\lambda\sin(k_yb-qx)\right]\right\}, 
\end{equation}
where 
\begin{equation}
q=\frac{ e b B\cos\theta}{\hbar} = \frac{\omega_B}{v_F} 
=\frac{\omega_0\cos\theta}{v_F}. 
\end{equation}
$\omega_B$ is the frequency at which electrons traverse the 
quasi-one-dimensional sheets of the Fermi surface~\cite{chaikin96} 
and
\begin{equation}
\lambda = \frac{2 t_y}{e b v_F B\cos\theta}  
\end{equation}
is the is the wavelength of the real space oscillations of the 
electron trajectories on the Fermi surface~\cite{chaikin96}. 
In a magnetic field the electron dispersion relation 
is {\em independent} of $t_y$, 
\begin{equation}
\epsilon_{\alpha}(k_x, k_y) = \alpha \hbar k_x v_F.
\end{equation}
All energies are relative to the Fermi energy. 
The GF can be calculated in a way similar to 
the one in Ref.~\onlinecite{moses99}, to give: 
\begin{equation}
G^{1+}_0(\rr,\rr',E) = -\frac{iL_x}{\hbar v_{F}}
\sum_{k_y, \alpha} \alpha
e^{i [ k_y (y-y') +
\alpha \lambda L ]}
e^{{\frac{i |x-x^{'}|}{\hbar v_{F}}}(E+i\Gamma)},
\label{GF1:eq}
\end{equation}
where
\begin{equation}
L = \sin(k_y b -q x^{'}) -\sin(k_y b - q x), \nonumber
\end{equation}
and $\Gamma$ is the electron scattering rate. 
The GF for the second layer differs by a gauge factor, 
$e^{\frac{ie}{\hbar}\Lambda(\rr-\rr')}$ and is
\begin{eqnarray}
G^{2+}_0(\rr,\rr',E)&=&-\frac{iL_x}{\hbar v_{F}}
\sum_{k_y, \alpha} \alpha
e^{i [ k_y (y-y') +
\alpha \lambda L ]}
\nonumber \\
&&\times e^{{\frac{i |x-x^{'}|}{\hbar v_{F}}}
\left(E+ev_FcB\sin\theta + i\Gamma\right)}.
\label{GF2:eq}
\end{eqnarray}

\subsection{Green function containing incoherence}
We now allow for the possibility that the motion in the interchain 
direction can be 
incoherent. The incoherence might come from polaron formation, strong 
electron-electron correlation, or any other many-body effect.  
In a formulation in terms of 
GFs a possible ansatz for the effect of incoherence is that the 
non-interacting GF is multiplied by a $y$-dependent factor 
\begin{equation}
G(\rr,\rr',\tau) = G_0(\rr,\rr',\tau) \sigma(y-y'), 
\label{incoh:eq}
\end{equation}
where $\sigma(y-y')$ depends on the process by which coherence is lost. 
The validity of this special form of GF can be seen 
for polarons in ,e.g., Refs.~\onlinecite{lundin03,kaye98,alexandrov}, and for 
electron-electron interaction, in ,e.g., Ref.~\onlinecite{lee96}. 
In a 2D strongly correlated model using the slave-boson approach~\cite{lee96} 
the 
electronic GF factorizes into $G=G_BG_F$, where $G_F$ is the free Fermion GF, 
and $G_B(\rr,\rr')=\exp\left(-|\rr-\rr'|^2/m_BT\right)$ is the bosonic 
GF containing the correlations, $m_B$ is the mass of the accompanying boson 
and $T$ the temperature.   
If the 2D-lattice is anisotropic (i.e., weakly coupled chains) 
the effect from the bosonic part will be even more pronounced. 
In a previous paper we studied transport in layered materials of 
polarons~\cite{lundin03}. For this case the GF contains two parts, 
one coherent, describing band motion of electrons weakly scattered by the 
phonons, and one incoherent, where localized polarons hop between sites. 
For the case of polarons Eq.~(\ref{incoh:eq}) is 
valid~\cite{alexandrov,lundin03}.  
Here, we do not specify the process responsible for the loss of coherence, 
but will just assume the general form given in Eq.~(\ref{incoh:eq}). 
The process involved in Eq.~(\ref{incoh:eq}) is the following.  
When the electron moves in the $(x,y)$-layer the $k_x$ momentum is conserved. 
Hence, there is no $x$-dependence in the term describing the 
incoherent contribution, $\sigma(y-y')$. Instead it describes the change in 
momentum in the $y$-direction as the particle jumps between  
$y$ and $y'$. The change in momentum is $\delta k_y$, which will be 
centered around zero so that for most of the time $k_y$ is unchanged. 
If the proposed form for the GF is correct, it could be visible in angle 
resolved photoemission spectra, which measures the spectral 
density~\cite{valla02,zwick97,sing03}. Later we will 
demonstrate that even a very small incoherent term gives rise to observable 
MA effects. 

\subsection{Interlayer conductivity}
Using the GFs, Eq.~(\ref{GF1:eq}) and Eq.~(\ref{GF2:eq}), 
in Eq.~(\ref{conduct:eq}), and the incoherence factor, $\sigma(y-y')$, 
we get a general expression for the conductivity: 
\begin{eqnarray}
&&\sigma_{zz}=
\nonumber \\ &&
\frac{e^2t_{\perp}^2c}{\hbar\pi L_xL_y}
\int {\mathrm d}\rr \int {\mathrm d}\rr'
\sum_{\alpha,k_y,k_{y'}} 
e^{i\left[ (k_y-k_{y'})(y-y')+\alpha\lambda(L+L')\right]}
\nonumber \\ && \times 
|\sigma(y-y')|^2
e^{-2\frac{|x-x'|}{\hbar v_F}\Gamma} \left( e^{iS}+e^{-iS} \right), 
\nonumber \\
\end{eqnarray}
where $S=\frac{ecB\sin\theta}{\hbar}|x-x'|$ is the change in gauge potential 
associated with interlayer transport. 
The summation over $\alpha$, the two Fermi sheets the electrons moves on, 
can be done and simplified 
This can be simplified to
\begin{eqnarray}
&&\frac{1}{2}\sum_{\alpha}e^{i\alpha\lambda(L+L')}=
\nonumber \\
&&2\cos\left\{4\lambda\sin[(k_y-k_{y'})b/2]\cos[(k_y+k_{y'})b/2]\right\} 
\nonumber \\
&&\times
\cos\left\{4\lambda\sin[(k_y-k_{y'})b/2]\cos[(k_y+k_{y'})b/2-
q/2(x-x')]\right\} \nonumber \\
&&+2\sin\left\{4\lambda\sin[(k_y-k_{y'})b/2]\cos[(k_y+k_{y'})b/2]\right\} 
\nonumber \\
&&\times
\sin\left\{4\lambda\sin[(k_y-k_{y'})b/2]\cos[(k_y+k_{y'})b/2-
q/2(x-x')]\right\}. \nonumber
\end{eqnarray}
Here, we introduce new variables, 
$k_y-k_{y'}=k_-$, $k_y+k_{y'}=k_+$, $x-x'=x_-$, $x+x'=x_+$, 
$y-y'=y_-$, and $y+y'=y_+$. 
We can then perform the integral over $x_+$ to give $L_x$, and the integral 
over $y_+$ to give $L_y$. 
We now use the representation of the trigonometric functions in terms of 
Bessel functions
\begin{eqnarray}
&&\cos[A\cos(k_+b/2-\Delta)]= \nonumber  \\
&&J_0[A]+2\sum_{k=1}^{\infty}(-1)^kJ_{2k}[A]\cos[2k(k_+b/2-\Delta)], 
\nonumber \\
&&\sin[A\cos(k_+b/2-\Delta)]= \nonumber  \\
&&2\sum_{k=0}^{\infty}(-1)^kJ_{2k+1}[A]\cos[(2k+1)(k_+b/2-\Delta)], 
\nonumber
\end{eqnarray}
where $J_l$ is a Bessel function of order $l$. 
The summation over $k_+$ can now be done by transforming it into an integral 
and we get 
\begin{eqnarray}
\sigma_{zz}&=&\frac{4 e^2t_{\perp}^2c}{\hbar b}
\int {\mathrm d}k_- \int {\mathrm d}x_- 
e^{-\frac{|x_-|}{\hbar v_F}\Gamma}
\cos\left(\frac{ecB\sin\theta}{\hbar}|x_-|\right) 
\nonumber \\ &\times&
\sum_{l=0}^{\infty} J_l\left[4\lambda \sin \left(\frac{k_-b}{2}\right)\right]^2
\cos(lqx_-) f(k_-),
\end{eqnarray}
where we introduced the distribution function 
\begin{equation}
f(k_-)=\int {\mathrm d} y_- e^{iy_-k_-}|\sigma(y_-)|^2,
\end{equation}
describing the spread (incoherence) in the (interchain) $y$-direction. 
The final step is the integration in $x_-$, which gives us the final 
expression 
\begin{eqnarray}
\sigma_{zz}(\theta)&=&\sigma_0
\sum_{l=-\infty}^{\infty}
\frac{\Gamma^2}{\Gamma^2+e^2v_F^2B^2(bl\cos\theta-c\sin\theta)^2}
\nonumber \\
&\times&\int {\mathrm d}k_- 
J_l\left[4\lambda \sin\left(\frac{k_-b}{2}\right)\right]^2
f(k_-)
, 
\label{cond:eq}
\end{eqnarray}
where we defined the conductivity in zero field, 
$\sigma_0\equiv\frac{8 t_{\perp}^2e^2 c}{v_F b \Gamma}$. 
Eq.~(\ref{cond:eq}) is the main result of this paper. 
This expression can be directly compared with those derived by other 
authors for alternate theories~\cite{maki92,osada02}.  
The MA appears as peaks in $\sigma_{zz}$ (dips in the MR), 
when the denominator has a minima. This will occur at angles when 
\begin{equation}
\tan\theta=\frac{b}{c}l, 
\label{MA:eq}
\end{equation}
i.e., at the MA. 

Recall that the function $f(k_-)$ indicates the amount of incoherence in 
the $y$-direction. If we have coherent particles in the 
$y$-direction, then, $k_y$ is always conserved so that $k_{y'}=k_y$, and  
the distribution will be  a delta function $f(k_-)=\delta(k_-)$. 
The sum over the Bessel functions collapses to only 
the $l=0$ term, and the result is 
\begin{equation}
\sigma_{zz}(B,\theta)=\sigma_0
       \frac{\Gamma^2}{\Gamma^2+\left(ecv_FB\sin\theta\right)^2}.
\label{condl0:eq}
\end{equation}
This agrees with the result from regular Boltzmann transport 
theory~\cite{blundell96}, and the MA effects are not seen.  

\begin{figure}[hbt]
\includegraphics[width=\columnwidth]{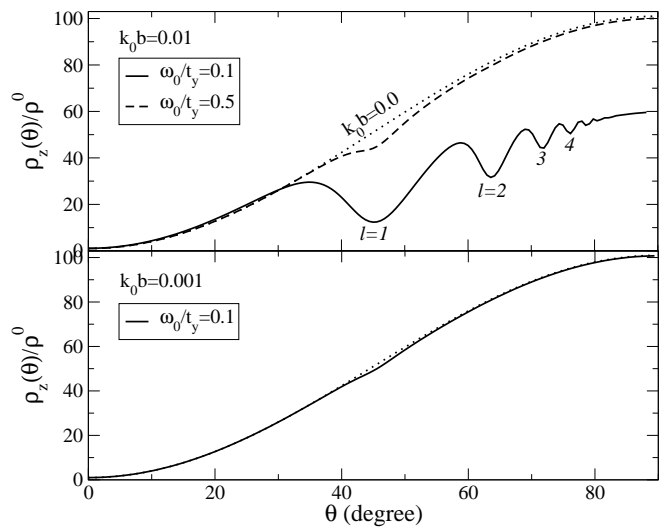}
\caption{\label{result:fig}
Interlayer magnetoresistance as a function of tilt angle, $\theta$, 
in the $y-z$-plane. 
Even for a very small incoherent hopping between the chains of molecules 
the magic angle effect are clearly seen.  
The parameter $k_0b$ is a dimensionless parameter describing spread in the 
distribution of momentum as the particle tunnels between the chains. 
$k_0b=0$ means full coherence, i.e., a delta-function distribution 
of $k_y$-values. 
The magnetic frequency, $\omega_0=ebv_FB$, is the frequency at which the 
electrons traverse the open sheets of the Fermi surface. 
$\rho^0$ is the resistivity in zero field, 
we used $b=c$ and $\omega_0/\Gamma=10$.
We also included, as a comparison, the result when no incoherence is present 
($k_0b=0$), given by Eq.~(\ref{condl0:eq}) in the text.} 
\end{figure}

If an incoherent term is present we will have some spread in $k_-$. 
To illustrate this 
we use $f(k_-)=\frac{1}{\sqrt{2\pi}k_0}e^{-\frac{(k_-)2}{2k_0^2}}$, 
meaning that the averaged momentum in the $y$-direction follow
\begin{equation}
\langle (k_y-k_{y'})^2 \rangle = k_0^2. 
\end{equation}
$f(k_-)$ has the property that 
it becomes a delta function if $k_0\rightarrow 0$, i.e., when the 
quasi-particles in the $y$-direction are coherent. 
The momentum in the $x$-direction is conserved, $k_{x'}=k_x$. 
We stress that the effects we are discussing are not sensitive to the 
particular form of $f(k_-)$ used, since it is an integrated quantity. 
$k_0b$ is a measure of how poorly the quasiparticle wave-vector is defined in 
the interchain direction. The electrons are coherent 
in the $y$-direction of the order of $k_0^{-1}$, meaning that if, say, 
$k_0b=0.01$, then the electrons are coherent on the order of 100 lattice 
constants in the $y$-direction. Thus, a value used below $k_0b=0.01$ still 
represents very well defined quasiparticles. 
A typical curve for the angular dependence of the interalyer 
magnetoresistance is shown in Fig.~\ref{result:fig}.
The value of the other parameters, $\frac{\omega_0}{t_y}$ and
$\frac{\omega_0}{\Gamma}$ are taken from typical experimental 
values. The decay, $\Gamma=\hbar/\tau$, comes from two experiments where the 
scattering time has been measured by magnetoresistance measurements and is 
$\tau=4.3$ps in Ref.~\onlinecite{danner94} 
[\TMTSFC at T=0.5K and ambient pressure] 
giving $\Gamma=0.15$meV, and $\tau=6.3$ps in Ref.~\onlinecite{lee98} 
[\TMTSF at T=0.32K and 8.2 kbar] 
giving $\Gamma=0.10$meV. The magnetic frequency, $\omega_0=ebv_FB$, is given 
by the Fermi velocity, $v_F=0.2$Mm/s in Ref.~\onlinecite{osada92}, and is 
equal to 1.08meV when the magnetic field 7T and 
$b$=7.711 \AA~\cite{danner94}. 
The hopping parameter in the $y$-direction, $t_y$ is given as 
31meV in Ref.~\onlinecite{lee98} and Ref.~\onlinecite{osada92}, but 
12meV in Ref.~\onlinecite{danner94}. 
In our numerical examples we use: 
$\frac{\omega_0}{\Gamma}=10$, $\frac{\omega_0}{t_y}=0.1$. 
The results are not that strongly dependent on the choice of these values, 
only the amplitude of the MA dips change. 
Here we have to point out that according to the experiments~\cite{danner94} 
there should be a dip when $\theta=90^{\circ}$, which is absent in our theory 
(see Fig.~\ref{result:fig}). This dip occurs 
when $B$ is parallel to the layers, and is therefore not a MA, and 
can not be described by our theory. As described in the introduction, 
it may be connected with the proximity to the superconducting 
state for the in-plane magnetic field~\cite{naughton97}. 

Note that by comparing our theory to the one by Osada~\cite{osada02} 
(which assumes non-interacting electrons) 
the incoherent term in the $y$-direction has a similar effect as a 
magnetic field in the $x$-direction. In particular we have 
\begin{equation}
B_x \leftrightarrow \frac{\hbar k_0}{ec}, 
\end{equation}
giving $B_x\sim 6.3$T if we use $k_0b=0.01$, . 
We see that even a very small incoherent part, $k_0b=0.01$, corresponds to a 
relatively large fluctuating field in the $x$-direction $B_x\sim 6.3T$. 
Thus, the larger the incoherence is (larger $k_0b$) the larger the 
corresponding effective field in the $x$-direction is, 
and the larger the MA dips 
in the MR are. This is consistent with the experimental result by Lee and 
Naughton~\cite{lee98}, where an increasing $x$-component of the magnetic field 
increased the size of the MR oscillations at the MA. 


\section{$\vec{B}$ in the $(x,z)$-plane}
\begin{figure}[hbt]
\includegraphics[width=\columnwidth]{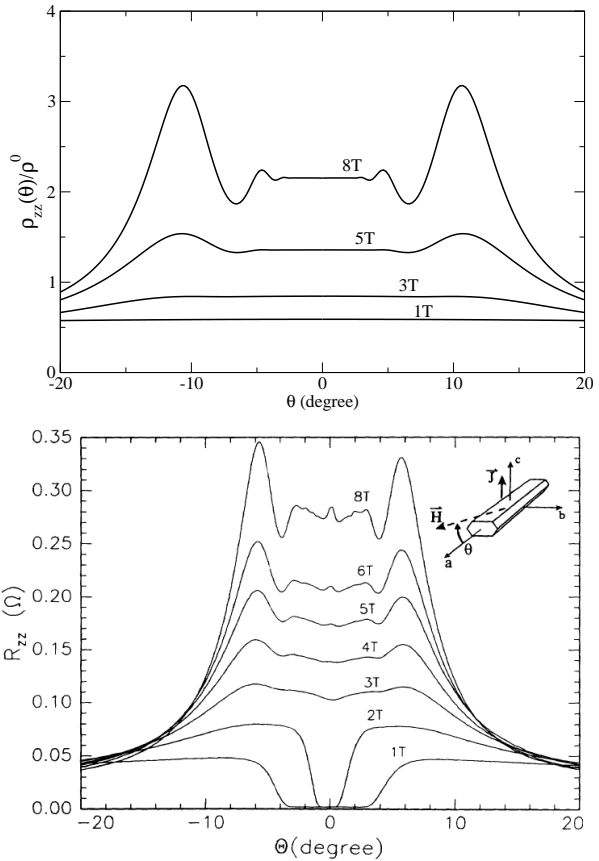}
\caption{\label{xz:fig}
Interlayer magnetoresistance as a function of the magnetic field 
direction in the $x-z$-plane. $\theta$ is the angle between the 
field and the $z$-axis. 
The upper panel shows a numerical calculation of the so called 
Danner-Kang-Chaikin oscillations~\cite{danner94}, 
from Eq.~(\ref{condxz:eq}) in the text. 
The theoretical curve can be compared with Fig.~1 from 
Ref.~\onlinecite{danner94} shown in the lower panel, with experiments done 
on (TMTSF)$_2$ClO$_4$ at ambient pressure and T=0.5K.  The dip around 
zero degree below 3T is due to the sample becoming superconducting. 
Note that $\theta$ denotes the angle between the magnetic field and the 
$x$-axis. 
We used $b=c$, $k_0b=0.001$, with $\omega_0/\Gamma=10$ and $\omega_0/t_y=0.1$ 
at $B=7T$. 
}
\end{figure}

If we instead apply the magnetic field in the $(x,z)$-plane the vector 
potential will be
\begin{equation}
\vec{A} = (0,xB_z-zB_x,0). \nonumber
\end{equation}
The derivation is very similar to the one presented above with the only 
difference that the gauge potential does not have any component depending on 
$|x-x'|$, but now depends on $y-y'$ instead. The result is that the integral 
over $x_-$ is simpler, but the integral over $y_-$ has an additional factor. 
This factor can be absorbed in 
the $y_-$ integral, the final result is 
\begin{eqnarray}
\sigma_{zz}(B,\theta)&=&\sigma_0
\sum_{l=-\infty}^{\infty}\int {\mathrm d}k_- 
J_l\left[4\lambda \sin\left(\frac{k_-b}{2}\right)\right]^2
\nonumber \\ &\times&
\frac{\Gamma^2}{\Gamma^2+(ev_FblB_z)^2}g(k_-), 
\label{condxz:eq}
\end{eqnarray}
where 
\begin{equation}
g(k_-)=\int {\mathrm d} y_- e^{iy_-\left(k_--\frac{ecB_x}{\hbar}\right)}
|\sigma(y_-)|^2=f\left(k_--\frac{ecB_x}{\hbar}\right). 
\end{equation}
The parameter $\lambda$ is $\frac{2t_y}{ebv_FB_z}$. 
The so called Danner-Kang-Chaikin oscillations~\cite{danner94} 
are observed  provided that 
\begin{equation}
\frac{ecB_x}{\hbar} \gg k_0,
\end{equation}
where $k_0b$ is the incoherence parameter. 
In Fig.~\ref{xz:fig} we compare the resulting resistivity ($1/\sigma_{zz}$) 
from Eq.~(\ref{condxz:eq}) with an experimental curve~\cite{danner94}. 
We did not adapt the parameters to the experiment, but just want to 
illustrate that this type of oscillations do appear in the theory presented. 
Note that we have used a smaller value for the incoherence parameter 
$k_0b=0.001$, compared to the value used in Fig.~\ref{result:fig}. 
This is justified by the fact that the experiment we compare with 
is performed for the ClO$_4$ compound and the 
oscillations in the $y-z$-plane are not as visible~\cite{naughton91} 
as for the PF$_6$ compound indicating a smaller incoherence factor. 

\section{$\vec{B}$ in the $(x,y,z)$-plane}
\begin{figure}[hbt]
\includegraphics[scale=1.0]{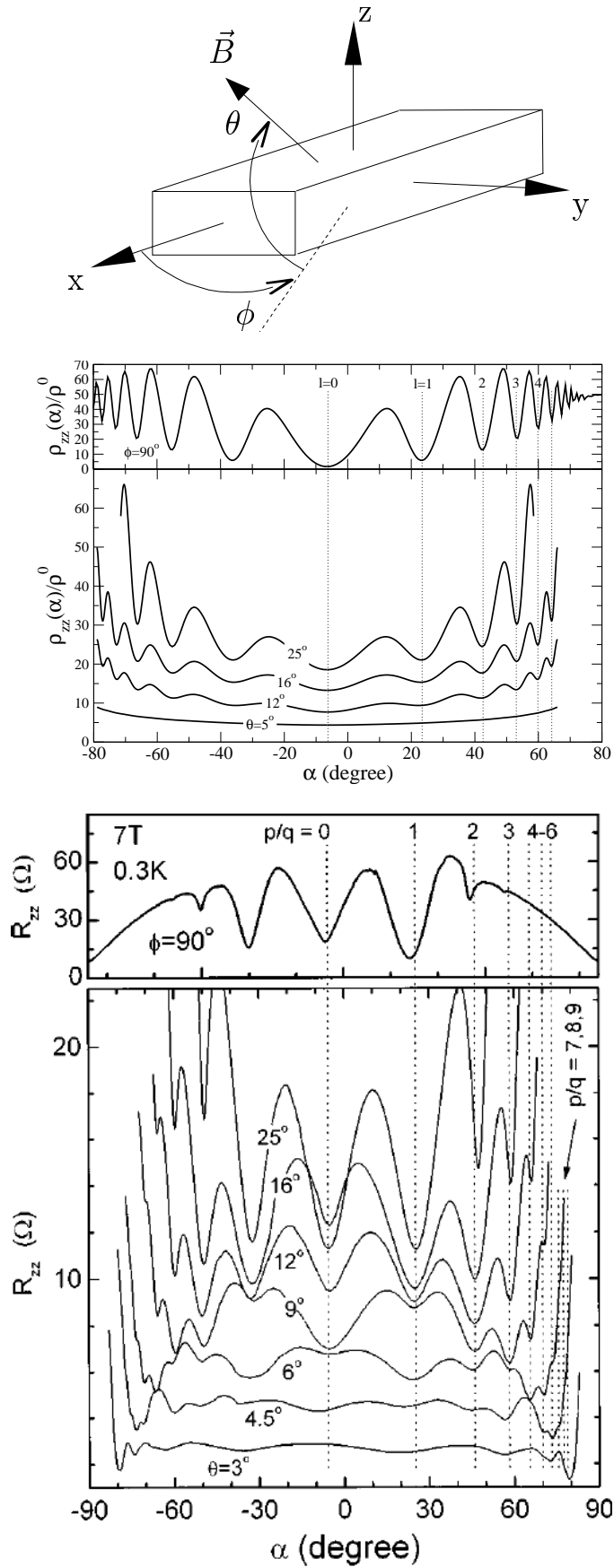}
\caption{\label{lee_calc:fig}
Interlayer magnetoresistivity versus $y-z$-plane angle, $\alpha$, defined via 
$\tan\alpha =\sin\phi /\tan\theta$ (see top figure). 
The middle panel shows the result from our numerical calculation of 
conductivity using Eq.~\ref{cond_all:eq}. 
Modulations appear at the magic angles as the angle $\alpha$ is increased. 
We used $b=7.581$\AA~and $c=13.264$\AA. 
The other parameters used are $k_0b=0.1$, $\omega_0/\Gamma=10$ and 
$\omega_0/t_y=0.1$.
The theoretical curve can be compared with Fig.~4 from 
Ref.~\onlinecite{lee98} shown in the lower panel. This is an experiment 
done at 0.32K on \TMTSF with an applied hydrostatic pressure of 
8.3 kbar to suppress the spin-density-wave state. 
}
\end{figure}

Combining the results from the calculations above we can get an expression 
for a field in a general direction, $(B_x,B_y,B_z)$. We get 
\begin{eqnarray}
\sigma_{zz}(\theta,\phi)&=&\sigma_0
\sum_{l=-\infty}^{\infty}
\frac{\Gamma}{\Gamma^2+e^2v_F^2(blB_z-cB_y)^2}\nonumber \\
&\times&\int {\mathrm d}k_- 
J_l\left[4\lambda \sin\left(\frac{k_-b}{2}\right)\right]^2
f\left(k_--\frac{ecB_x}{\hbar}\right), \nonumber \\
\label{cond_all:eq}
\end{eqnarray}
note that $\lambda$ is a function of $B_z$. 
In Fig.~\ref{lee_calc:fig} we compare results from this expression with 
the experimental results of Lee and Naughton~\cite{lee98}, by 
identifying the angles defined in Fig.~\ref{lee_calc:fig}, 
as follows, 
\begin{equation}
\left\{
 \begin{array}{l}
 B_x=B\cos\theta\cos\phi \\
 B_y=B\cos\theta\sin\phi \\
 B_z=B\sin\theta
 \end{array}
 \right.
  ,
\end{equation}
where the definition of $\theta$ and $\phi$ follows 
Ref.~\onlinecite{lee98}, (see the upper panel in Fig.~\ref{lee_calc:fig}). 
As the angle $\theta$ between the $(x,y)$-plane and the direction of the field 
is increased, the oscillations start to appear. 
The similarities to Fig. 4 in Ref.~\onlinecite{lee98} are striking. 
Changing the parameters in the model does not change the general features 
of this plot. 

\section{Discussion}
In summary we have presented an explanation in terms of many-body effects 
of the appearance of magic angle effects in the interlayer magnetoresistance. 
The MA appears naturally from, even a small, incoherent 
contribution to the inter-chain hopping. The hopping in the 
most conducting direction is assumed to be coherent, and in the least 
conducting direction incoherent. Momentum can change in the direction 
between the one-dimensional chain of molecules. 
This is described by a distribution 
function which is centered around zero, letting most quasi-particles 
retain their momentum when hopping. 
We used an explicit form of the interlayer Green function, which 
can be directly observed in a angle resolved photoemission spectra. 
Unlike present explanations~\cite{osada92,chashechkina01,blundell96}, 
the theory does not assume any long distance hopping between non-adjacent 
quasi-one-dimensional molecules in different layers, 
where the overlap is quite small, 
only a nearest neighbor interlayer overlap. 
The shape of the Fermi surface is not affected by the incoherence. 
Numerical calculations produce results similar to experimental results. 

\acknowledgments
This work was supported by the Australian Research Council.
U.\ Lundin acknowledges the support from the Swedish foundation for
international cooperation in research and higher education (STINT).
We thank P.M.\ Chaikin for helpful discussions. 

\bibliography{paper}

\end{document}